# Adoption of ICT innovations in the agriculture sector in Africa: A Systematic Literature Review

Claudia Ayim[a], Ayalew Kassahun[b], Bedir Tekinerdogan[c], Chris Addison[d]

Wageningen University, Information Technology, Wageningen, The Netherlands

[a] claudia.ayim@wur.nl, [b]ayalew.kassahun@wur.nl, [c]bedir.tekinerdogan@wur.nl, [d]Addison@cta.int

**Abstract**

According to the latest World Economic Forum report, about 70% of the African population depends on agriculture for their livelihood. This makes agriculture a critical sector within the African continent. Nonetheless, agricultural productivity is low and food insecurity is still a challenge. This has in recent years led to several initiatives in using ICT (Information Communication Technology) to improve agriculture productivity. This study aims to explore ICT innovations in the agriculture sector of Africa. To achieve this, we conducted a SLR (Systematic Literature Review) of the literature published since 2010. Our search yielded 779 papers, of which 23 papers were selected for a detailed analysis following a detailed exclusion and quality assessment criteria. The analysis of the selected papers shows that the main ICT technologies adopted are text and voice-based services targeting mobile phones. The analysis also shows that radios are still widely used in disseminating agriculture information to rural farmers, while computers are mainly used by researchers. Though the mobile-based services aimed at improving access to accurate and timely agriculture information, the literature reviews indicate that the adoption of the services is constrained by poor technological infrastructure, inappropriate ICT policies and low capacity levels of users, especially farmers, to using the technologies. The findings further indicate that literature on an appropriate theoretical framework for guiding ICT innovations is lacking.

## 1  Introduction

The agriculture sector in Africa is less developed and food insecurity is still a challenge. Though the continent has enormous natural resources and the agricultural potential is high, many countries are still net importers of food. An African Development Bank report (Moyo, Bah, & Verdier-Chouchane, 2015) revealed that the continent imports about US$25 billion of food crops annually. According to the authors, the level of value addition and processing of agricultural commodities is also low and post-harvest losses are high (in sub-Saharan Africa alone averaging about 30 percent of total production). Agriculture, however, remains a significant sector within the continent. It is the main source of income for the majority of its rural people. The agriculture sector accounts for almost two-thirds of the total employment and about 75% of domestic trade (World Bank, 2008). With the majority of the rural population depending on agriculture for their livelihoods, the growth and development of the sector are critical.

Growth and development of the agriculture sector can be achieved through the effective deployment of Information Communication Technology (ICT). According to the Food and Agriculture Organisation (FAO, 2017), ICT has been a significant contributor to the growth and socio-economic development in countries and sectors where they are well deployed. The effective integration of ICT in the agriculture sector in American and European countries has led to tremendous improvement in agriculture value chain efficiency and productivity. For instance, traceability technologies such as blockchain, and radio frequency identification (RFID) has enabled transparency and efficiency throughout the food chain through tracking and tracing of food from farm to fork. This makes it possible to identify the source of any food-related incident in case a food safety issue occurs. However, such transformation has yet to take place in Africa.

In recent years, efforts to transform the sector has led to the propagation of several mobile-based applications and services. A recent digitalisation report by the Technical Centre for Agriculture and Rural Cooperation (Digitalisation of African Agriculture, 2019) revealed that 33 million smallholder farmers are currently reached by digital applications as of 2019 and this is projected to reach 200 million by 2030. These applications are diversified targeting advisory and information services, market linkages, financial access, and supply chain management, with advisory and information service dominating the market (Digitalisation of African Agriculture, 2019). El Bilali & Allahyari (2018) assert that ICT-based innovations can improve rural livelihoods and empower smallholder farmers in developing counties by enhancing their connectivity and increasing access to accurate and timely agriculture information. For example, Esoko which is a technology platform in most African countries uses a combination of mobile and web services to improve access to extension services and market information. This reduces the costs of searching for market information and provide real-time weather and extension advice to farmers.

Innovative ICT's ranging from computers, radio, television and mobile phones to advanced technologies such as blockchain, artificial intelligence, cloud computing, Internet of Things (IoT) and big data analytics are among the current trends (OECD, 2017). El Bilali & Allahyari (2018) argue that these disruptive ICT trends hold the potential to contribute to sustainability transitions in agriculture by increasing efficiency, enhancing transparency and traceability. Iliyas (2014) stressed that remote sensing using satellite technologies, and geographical information systems can be used to increase agricultural output. Furthermore, big data analytics can be used to provide predictive insights in farming operations, drive real-time operational decisions, and redesign business processes (Wolfert et al., 2017)(Ahoa et al, 2020)(Kassahun et. al, 2020). Precision agriculture involving the use of several technologies such as Global Positioning System (GPS), Geographic Information Systems (GIS), mobile computing, advanced information processing, and software can be used for comprehensive data on production variability in both space and time (Zhang et al, 2002)(Koksal & Tekinerdogan, 2019)(Verdouw et al., 2019). With ICT recognized as a significant contributor to the growth and development of agriculture, its application in recent years has gained increasing attention in many developing countries.

There are however few literature done on ICT innovations in the past. Zewge and Dittrich (2017) conducted a systematic mapping to describe the state-of-the-art of ICT for Agriculture research in developing countries. Their work presented journal and conference papers published between 2006 to 2014 in developing countries. They found 846 papers and considered 57 papers for qualitative data synthesis. Their study identified mobile phones, computers, telecenters, and the internet as the main ICTs in developing countries with the mobile phone being the preferred technology, especially in rural areas. The study states that appropriate design solutions that take social, as well as technical issues into account, are still scarce. Similarly, Lwoga and Sangeda (2019), conducted a systematic review of reviews to describe ICT for development research trends, methodologies and conceptual frameworks in developing countries. The authors considered 8 papers from 100 originally selected academic journals, conference papers, books, and working papers published between January 1990 and July 2017. The study identified qualitative data synthesis as the main methodology used in the selected reviews. Also, it was evident from the conceptual frameworks that there is limited use of user-centered design research in the development of ICT applications. The study, therefore, recommends a paradigm shift from developing technologies for users to designing and developing applications with users to gain insights and enhance collective problem definition in the given context.

We are however unable to find a systematic literature review on the topic of ICT innovations in Africa. This study, therefore, aims, to assess the current state of adoption of ICTs specifically within the agriculture domain in Africa. To achieve this, we performed a Systematic Literature Review (SLR) to make a comprehensive and rigorous review of the existing literature. The review of literature on ICT innovation within the continent is valuable to contribute to the body of knowledge within the continent.

The paper is organized as follows: the introduction is discussed in section 1, section 2 presents the research methodology, section 3 presents the results of the SLR, section 4 presents the discussion of the findings, and the conclusion is presented in section 7.

## 2 Research Methodology

The review protocol proposed by Kitchenham et al., (2019) was followed in this study (see **Figure 1**). Following this protocol, we started by identifying the research questions the SLR has to address. This was followed by detailing the search strategy, including defining study selection criteria. Then we defined quality assessment criteria in the form of a well-defined checklist to assess the selected studies. We then applied the selection and quality criteria to select primary studies. We then extracted the relevant review data from the primary studies and analysed the data.

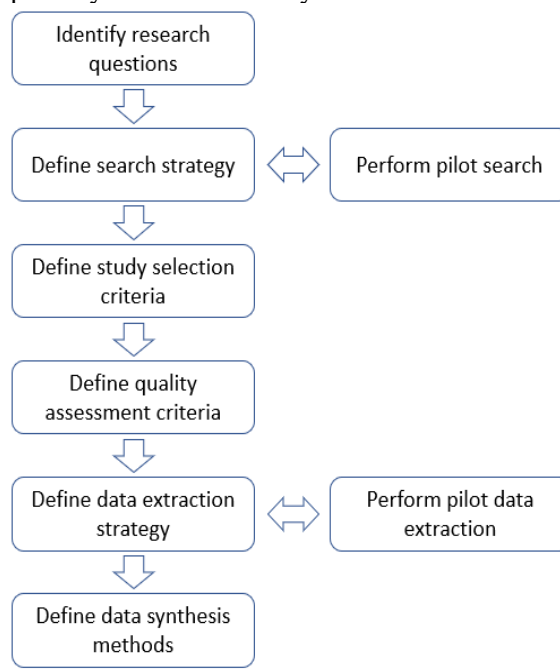

**Figure 1**: Review protocol used in our study. Adapted from (Kitchenham et al., 2009)

### 2.1 Research questions

In this paper, we are interested in investigating empirical studies on the current state of adoption of ICT innovations within the agriculture sector in Africa. To achieve this objective, the following research questions were defined:
RQ1. What are the main ICT technologies used within the agriculture domain?
RQ2. Whom were the agriculture stakeholders identified in the studies?
RQ3. Which African countries have applied the ICT technologies?
RQ4. What are the agricultural domains considered for ICT innovation?
RQ5. What are the frameworks used in the studies?
RQ6. What are the challenges in ICT adoption?

### 2.2 Search strategy

The search strategy was defined by explaining the search scope, search method, and search string. The search scope consisted of the publication venue and the time of publication. In terms of publication time, the studies focused on papers published from 2010 to February 2019, however, this was specified at the inclusion and exclusion criteria stage. The following well-known databases were used to search for the targeted research papers: Scopus, ScienceDirect, IBI/Inform and Wiley online. For the search methods, an automatic and manual search was used. The automatic search was achieved by searching electronic databases using defined search strings. The manual search was achieved by manually searching suggested papers or papers referenced by authors that were not identified through automatic search. The research questions were decomposed into keywords along with a list of synonyms before pilot searching to get relevant studies. The keywords are grouped in a search string as follows:

(*ICT OR digitalization OR digitalisation*) AND (*Africa OR "Sub-Saharan Africa"*) AND (*agriculture OR "agri-food"*) AND (*"innovation model" OR "innovation framework"*).  We searched the following databases using the search string: *Scopus* and *ScienceDirect*. For *Wiley online* and *ABI/INFORM,* the search string was slightly updated by adding *farming* in the third group of keywords to retrieve relevant papers.

The result of the overall search process after applying the search queries is given in **Table 1**.

**Table 1**: Overview of literature search results

| Source | Retrieved[1] | Included[2] | Selected[3] | Method |
|---|---|---|---|---|
| Scopus | 106 | 12 | 7 | Automatic |
| ScienceDirect | 132 | 5 | 4 | Automatic |
| Wiley Online | 206 | 4 | 5 | Automatic |
| ABI/INFORM | 322 | 4 | 4 | Automatic |
| Other Channels | 13 | 3 | 3 | Manual |
| Total | 779 | 28 | 23 | |

[1] Papers retrieved through automated and manual search
[2] Papers remaining after applying the exclusion criteria
[3] Papers remaining after applying both the exclusion and the quality assessment criteria

### 2.3 Study selection criteria

We detailed our exclusion criteria to serve as a way of reducing the number of studies while retaining relevant studies of interest. A set of criteria was applied to the identified papers to exclude the papers that do not correspond to the purpose of the study. The exclusion criteria used in the study are presented in **Table 2**. The exclusion criteria were applied to the 779 papers obtained from the automatic and manual search. This was done by firstly reading the title and abstract of papers and secondly looking at the paper as a whole.  After applying the criteria, 28 papers were kept for further assessment.

**Table 2**: Selection criteria

| No. | Criteria |
|---|---|
| SC1 | Papers published before 2010 |
| SC2 | Papers not written in English |
| SC3 | Papers without full text |
| SC4 | Papers do not relate to the agriculture domain |
| SC5 | The abstract does not discuss any ICT innovations and/framework |
| SC6 | Duplicate publication from multiple sources |

### 2.4 Quality assessment criteria

The remaining 28 papers were further assessed according to a well-defined quality checklist presented in **Table 3**

**Table 3**. This was done to provide a more detailed exclusion criterion. The quality assessment instruments used in the studies were based on (Kitchenham et al., 2009) criteria.  This criterion was divided into four main categories based on the factors that could bias the results namely: reporting, relevance, rigor, and credibility. Firstly, the quality of the reporting was analysed based on the aim, clarity, and coherence of the studies. Then, the issue of rigor was judged based on the extent to which the studies provide value for research and practice. The relevance of the studies was also assessed according to how thorough and complete all the aspects that the paper promised to answer were answered.  Finally, the credibility is assessed according to the extent to which the findings and the conclusions of the studies are meaningful and logical. The answers to the quality checklist questions were deployed on a numerical scale numbered with 0 for "no", 0.5 for "somewhat" and 1 for "yes" with regards to how well the paper answers the questions asked. After reading the full text of the 28 papers and applying the quality checklist, 23 papers were extracted based on their good quality scores. The detailed scores of the quality checklist are presented in Appendix B: Quality Assessment Checklist.

Table 3. Assessment criteria for identified primary studies

| No. | Question |
|---|---|
| Q1 | Is the aim of the study clearly stated? |
| Q2 | Is the scope and context of the research clear? |
| Q3 | Is the reporting clear and coherent? |
| Q4 | Are the theories used clear? |
| Q5 | Is the research methodology well presented? |
| Q6 | Are all study questions answered? |
| Q7 | Is the research process adequately documented? |
| Q8 | Is there a comprehensive description of ICT innovation and/frameworks? |
| Q9 | Does the conclusion relate to the aim of the study? |
| Q10 | Are the limitations of the research clearly stated? |

## 2.5 Data extraction

At this stage, a data extraction form was developed to accurately extract data from the primary studies. Pilot data extraction was performed and all the fields relevant for addressing our research questions were agreed upon. The data extraction form is available in Appendix C: Data Extraction Form. This form contains 15 elements, which include standard information such as authors, title, publication year, document type and data repository. It also contains elements needed for answering the research questions like the considered ICT domain, the considered agriculture domain, and the challenges in ICT adoption. A record of the extracted information was kept in a spreadsheet to support the process of synthesizing the extracted data.

## 2.6 Data synthesis

The purpose of the data synthesis is to summarize and present the findings of the primary studies in a manner suitable for answering our research questions. Based on the research objective and findings from the primary studies that were selected, this paper fits in a qualitative study and hence a descriptive synthesis of the extracted data was performed. We analysed the individual studies and the set of studies as a whole. Studies with similar or same meaning but different concepts were identified and grouped under one concept. For the challenges in the adoption of ICT's, we analysed and grouped them into four main concepts.

## 3 Results

This section presents the findings from the primary studies. Firstly, relevant descriptive statistics of the 23 selected papers are provided. The results corresponding to our research questions are also presented in this section.

## 3.1 Overview of selected studies

As previously stated, the review spelled out the time boundary of the search to include papers from 2010 to 2019 which was defined at the study selection stage. The year-wise distribution can be seen in **Figure 2**

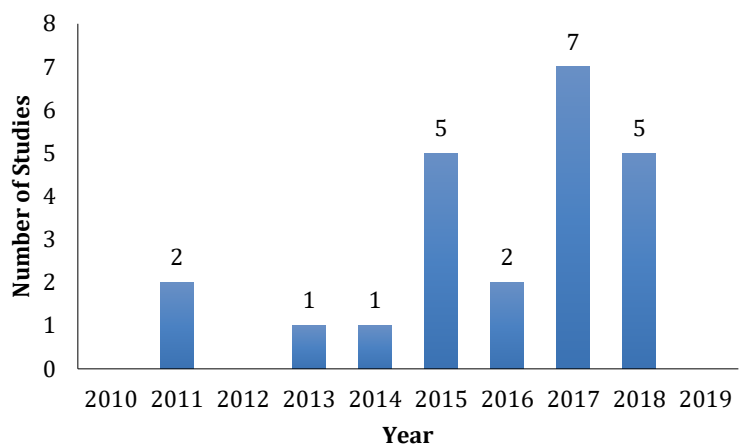

**Figure 2**: Year-Wise distribution of the 23 primary studies

Concerning the publication channels, diverse journals were retrieved. While some journals were broad in terms of the country scope, others were context-specific. The "Electronic Journal of Information Systems in Developing Countries" is an example of a context-specific journal that looks for publication only in developing economies. Also, the highest number of primary studies (6) were retrieved from the "Electronic Journal of Information Systems in Developing Countries". The second most popular channel is "Information Technology for Development" with 3 primary studies (see **Table 4**).

**Table 4**: Publication channel and the number of occurrences of the selected studies

| *Publication source* | # |
|---|---|
| Progress in Development Studies | 1 |
| Telecommunications Policy | 1 |
| Electronic Journal of Information Systems in Developing Countries | 6 |
| Information Technology for Development | 3 |
| Journal of Agricultural Education and Extension | 1 |
| Journal of Enterprise Information Management | 1 |
| Technological Forecasting & Social Change | 1 |
| South African Journal of Information Management | 1 |
| Computers and Electronics in Agriculture | 1 |
| Information Processing in Agriculture | 1 |
| African Journal of Agricultural Research | 1 |
| International Journal of Agricultural Sustainability | 1 |
| Journal of Rural Social Sciences | 1 |
| Society and Business Review | 1 |
| Journal of Agricultural & Food Information | 1 |
| Journal of Agricultural Informatics | 1 |

### *3.2 Methodological quality*

This section presents the quality of selected 23 studies. As already specified the quality assessment criterion was divided into four main parts namely: reporting, relevance, rigor, and credibility. The answers to the quality checklist questions were deployed on a numerical scale numbered with 0 for "no", 0.5 for "somewhat" and 1 for "yes" with regards to how well the paper answers the questions asked. The overall methodological quality scores are summarized in **Figure 3**. All of the criteria were taken into account, and 13 of the studies (57%) have scores greater than or equal to 8 which can be said to be of good quality while 10 (43%) of the remaining primary studies with scores less than 8 is of medium quality.

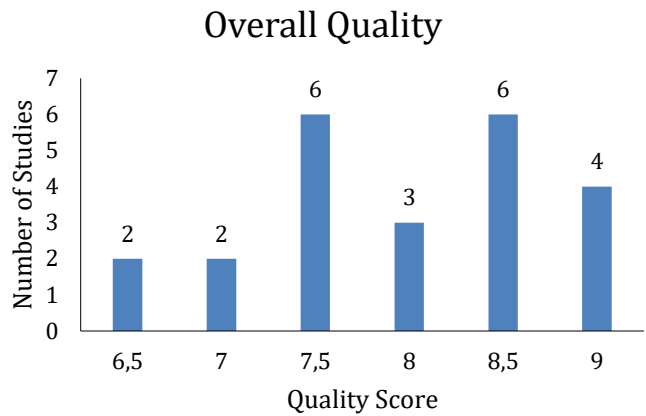

**Figure 3**: Overall quality of the 23 selected studies

*3.3 What are the main ICT technologies used within the agriculture domain? (RQ1)*

The purpose was to identify the main ICT technologies used within the agriculture domain as presented in **Table 5**

**Table 1**. The majority of the studies identified mobile phones as the widely used ICT tool within the agriculture sector. According to the studies, the proliferation of mobile phones within the African continent has led to the development of mobile-based applications and services within the sector. These services and applications are mostly targeted at farmers and range from the provision of farmers with agriculture information such as market prices of farm produce, weather, agriculture input, and improved agriculture techniques. The study [16] cited Ken Call which is a farmer's helpline service in Kenya that provides agricultural advice and information to support smallholder farmers. Farmers are provided with information on improved agricultural production, inputs, processing, climate and market information through the medium of a mobile phone. For the usage of this service, farmers call the helpline with specific questions that are addressed by agricultural experts and subject matter specialists.

Studies [2][4][6][8][15][16][19][22] also highlighted the importance of radio in disseminating agriculture information to rural farmers. According to study [2], the radio remains the most widely used medium in rural Africa. The study cited an example of an interactive radio project that was designed to help small-scale farmers to increase their production in Tanzania, Uganda, Malawi, and Ethiopia. The radio programme included regular radio broadcasts on agricultural information to farmers. The programme allowed farmers to ask questions through SMS or by calling and the responses are disseminated via the radio. Studies [2][4][6] explained that the use of radio by most farmers is because the radio programmes are broadcasted on community radios and in the local language of the farmers making it easy for farmers to understand the content of the information.

According to studies [5][7][9][15][16][23], ICT technologies such as computers and remote sensing technologies are mostly used by and researchers and agribusiness experts. These technologies were however identified to be inaccessible to most farmers because of lack of skills and the financial means to own such technologies.

Table 5: ICT technologies identified from primary studies

| ICT tools | Studies |
|---|---|
| Mobile phone | [1][2][3][4][5][6][8] [10] [12] [13] [14] [15] [16] [18] [19] [21] [22][23] |
| Radio | [2][4][6][8] [15] [16] [19][22] |
| Television | [6] [15] [16] [19][22] |
| Computer | [5][7][9] [15] |
| Remote sensing | [9][16][23] |

### 3.4 Whom were the agriculture stakeholders identified in the studies? (RQ2)

This research question aims to identify the agriculture stakeholders under investigation in the selected primary studies (see **Table 6**). A considerable number of studies investigated the use of ICTs among farmers. According to the studies, farmers use ICTs such as mobile phones for contacting extension workers, accessing prices of agriculture inputs and commodities. Four of the primary studies (5, 7, 9 and 20) discuss the use of ICTs by researchers, extension workers, and agribusinesses. Here the focus is on the use of ICTs in agriculture research. Study 5 examines the access and utilization of ICTs among researchers and extension workers. Study 7 and 20 discuss social factors that contribute to the adoption of ICTs among agribusinesses operating in rural areas. In study 9, socio-technical factors that limit the usage of ICTs by agriculture researchers are discussed.

Table 6: Identified agriculture stakeholders

| Study | Agriculture Stakeholders | | | |
|---|---|---|---|---|
|  | Farmers | Researchers | Extension workers | Agribusinesses |
| 1 | X |  |  |  |
| 2 | X |  |  |  |
| 3 | X |  |  |  |
| 4 | X | X |  |  |
| 5 |  | X | X |  |
| 6 | X |  |  |  |
| 7 |  |  |  | X |
| 8 | X |  |  |  |
| 9 |  | X |  |  |
| 10 | X |  |  |  |
| 11 | X |  |  |  |
| 12 | X |  |  |  |
| 13 | X |  |  |  |
| 14 | X |  |  |  |
| 15 | X | X | X |  |
| 16 | X |  |  |  |
| 17 | X |  |  |  |
| 18 | X |  |  |  |
| 19 | X |  |  |  |
| 20 |  |  |  | X |
| 21 | X |  |  |  |
| 22 | X |  |  |  |
| 23 | X |  |  |  |
| **Total** | **19** | **4** | **2** | **2** |

### 3.5 Which African countries have applied the ICT technologies? (RQ3)

Classifying the primary studies based on geographical location within which the studies were conducted, it was evident that the research was undertaken in 12 different countries namely: Ghana, Ethiopia, Mali, Tanzania, Zimbabwe, Kenya, Nigeria, Uganda, South Africa, Malawi, Mozambique, and Burkina Faso (see **Figure 4**).

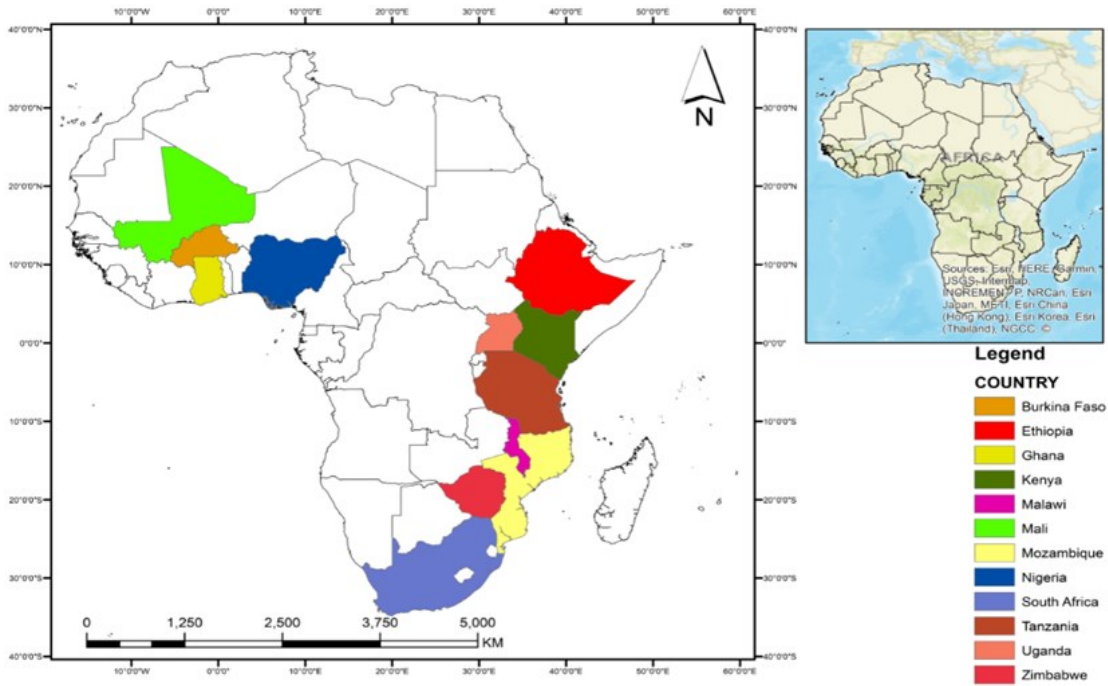

**Figure 4**: Distribution of primary studies by location

*3.6 What are the agricultural domains considered for ICT innovation? (RQ4)*

To answer this research question, the targeted agriculture domain of the 23 selected primary studies were analysed (see **Figure 5**). The majority of the primary studies (52%) emphasized the agriculture sector in general in studying the ICT innovations within the sector. 39% of the studies emphasized the crop sub-domain while the focus on the livestock and agroforestry sub-domain constituted 4% and 5% respectively of the primary studies.

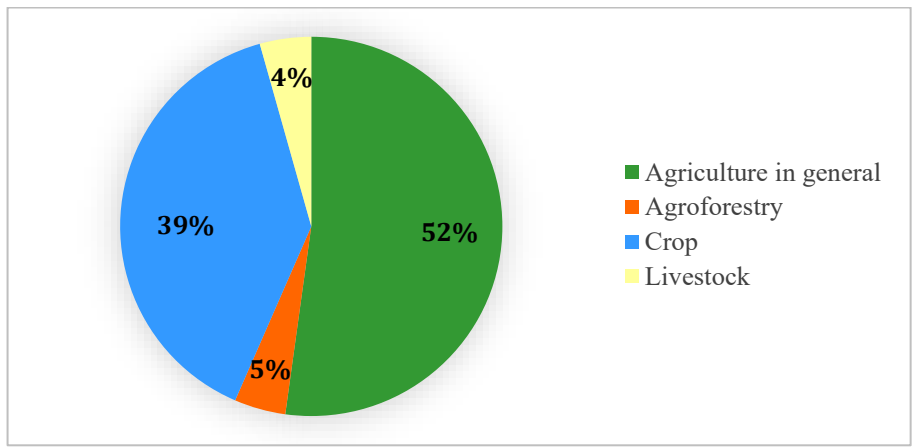

**Figure 5**: The considered agriculture domain of the primary studies

*3.7 What are the frameworks used in the studies? (RQ6)*

This research questions aimed to identify the various frameworks used in the studies. The study identified four main theoretical frameworks from the primary studies as presented in **Table 7Table 7**.

Table 7: Theoretical Frameworks identified from SLR

| Type of theory | Construct | Countries | Studies |
|---|---|---|---|
| Technology acceptance model (TAM) | • Perceived usefulness<br>• Perceived ease of use | Uganda, South Africa, Nigeria | [7][10][11] [17][20] |
| Diffusion of innovation theory (DO1) | • Relative advantage<br>• Compatibility<br>• Simplicity<br>• Observability<br>• Trialability | Zimbabwe, Mali, Kenya, Nigeria | [3][5][6][7] [11][12][14] [17][20] |
| Social network theory (SNT) | • Social influence<br>• Network externalities | Nigeria | [7] [20] |
| The unified theory of acceptance and use of technology (UTAUT) | • Performance expectancy<br>• Effort expectancy<br>• Facilitating conditions<br>• Social influence | Ethiopia | [12] |

Here we elaborate on the results in **Table 7**. From the SLR we identified four theoretical frameworks. These are the technology acceptance model, the diffusion of innovation theory, the social network theory, and the unified theory of acceptance and use of technology.

The diffusion of innovation theory (DOI) and the technology acceptance model (TAM) were the dominant theories used in the selected studies. For instance, studies [11] [17] used the DOI and TAM in providing a conceptual foundation for their studies. Study [10] applied the TAM to investigate the factors that determine the adoption of mobile phones among Uganda farmers. Similarly, study [11] used TAM and DOI to explore the adoption of ICTs by small-holder farmers. The TAM has two main constructs, the perceived usefulness (PU) and the perceived ease of use (PEOU) which are the predictors of the intention of a user to use technology. PU denotes the extent of an individual's belief that using technology would enhance his or her performance while PEOU is the measure of the extent to which an individual believes that using technology would require less effort. The DOI postulates five main characteristics of innovation which determine the rate of adoption namely: relative advantage, compatibility, complexity, observability, and trialability. Relative advantage is the degree to which an innovation is perceived as been useful. Compatibility talks about the degree to which an innovation is perceived as consistent with the existing values and needs of potential users. Complexity is the degree to which an innovation is perceived as relatively less complex and easy to understand and use. Observability on the hand is the degree to which the results of an innovation are visible to others while trialability describes how easily the innovation can be explored.

Studies [7] and [20] used the Social network theory (SNT) to identify the social imperatives that impact the adoption of ICTs by agribusinesses in Africa. The SNT refers to the social structure of relationships and links between individuals, businesses, and organizations within a particular context that can influence ICT adoption behaviour.

Study [12] adopted the UTAUT in identifying the drivers of mobile SMS adoption by farmers. The UTAUT incorporates social influence in determining behavioural intention. The UTAUT postulates that behavioural intentions are determined by four key constructs: performance expectancy, effort expectancy, social influence, and facilitating conditions. Performance expectancy is the degree to which using technology will provide benefits to users in performing certain activities. Effort expectancy is the degree of ease associated with farmers' use of technology. The underlying concept of social influence is that individuals can be influenced by the perceived social pressure of important others. Facilitating conditions are also perceived enablers or barriers in the environment that influence a person's perception of ease or difficulty of performing a task as identified in the studies.

### 3.8 What are the challenges in ICT adoption (RQ7)?

**Table 8**: Identified challenges from primary studies

| Challenges | Description of challenges | Studies |
|---|---|---|
| Poor infrastructure | • Unreliable electricity supply<br>• Network connectivity problem | [4][5][9][16][21] [22[23] |
| Low capacity | • Low illiteracy rate<br>• Lack of ICT knowhow<br>• Poverty | [4][5][9][16][21][22][23] |
| Poor ICT policies | • Lack of appropriate ICT policies<br>• Poor monitoring of ICT projects<br>• Poor funding of ICT initiatives | [4][5][9][16][21][22][23] |
| Inefficiencies in agriculture institutions | • Weak network between agriculture institutions<br>• Poor incentives to motivate stakeholders | [9][23] |

The identified challenges from the primary studies affecting the adoption of ICT's within the agriculture sector as presented in **Table 8** is explained below.

**Poor infrastructure**

Study 4, 5, 9, 16, 21, 22, and 23 discuss poor infrastructure as a challenge to the adoption of ICT. According to the studies, infrastructure development is still in its infancy in most African countries. This is however prevalent in most rural communities. Most rural communities in Africa is characterized by poor road network, no access to electricity, and poor network connectivity. With the majority of the farming population living in rural communities, the absence of these technological infrastructure poses a barrier to ICT adoption.

**Low capacity**

The adoption of ICT is constrained by poverty, lack of ICT knowhow and illiteracy as discussed in studies 4, 5, 9, 16, 21, 22, and 23. One important aspect identified in the studies is the illiteracy rate among farmers. Illiteracy rate is very high amongst smallholder farmers which affects their ability to effectively understand and manage the use of ICT tools. Also, farmers are unable to afford the cost of servicing mobile phones and paying for extension advisory services rendered by ICT innovators because of their low standard of living. Also, studies 16 and 23 highlights the role of gender in ICT adoption. According to the studies, women are not equally able to access and use ICTs due to their unequal access to opportunities such as income and education.

**Poor ICT policies**

In study 4, 5, 9,16,21,22, and 23 poor ICT policies is discussed. Agriculture development in most African countries has been undermined by poor policies that constrain market entry and the effective allocation of resources. According to the studies, poor prioritization of ICT initiatives, weak monitoring, poor implementation and integration of ICTs within the sector are due to poor ICT policies. This affects the adoption of ICT in agriculture, especially those targeting rural communities and rural development.

Also, the adoption of ICT is constrained by uncoordinated and scattered ICT initiatives as discussed in the studies.

**Inefficiencies in agriculture institutions**

According to study 9 and 23, ICT adoption is hindered by inefficiency in agricultural research institutions in Africa. Diversity among agriculture stakeholders, lack of commitment and accountability of policymakers and agriculture experts, and lack of incentives to undertake ground-breaking ICT research affect the effective adoption of ICTs. Also, weak institutions, lack of information sharing, and lack of awareness of existing ICT facilities and resources affect ICT adoption as discussed in studies 9 and 23.

## 4 Discussion

This systematic review presented the findings of the 23 selected primary studies on the adoption of ICT's within the agriculture domain in Africa. To the best of our knowledge, this is the first SLR on the-state-of-art of ICT innovations in Africa. We could identify that over the past years, high-quality papers have been published on the adoption of ICT's. A qualitative research approach was employed in summarizing the main findings from the selected papers. The analysis of the primary studies showed the proliferation of ICTs such as television, radio, computer, and mobile phone in the agriculture sector. Mobile phone was identified as the predominant ICT used within the sector which is in line with the findings of Zewge and Dittrich (2017). Apps and services on mobile phones which are mostly targeted at farmers allow farmers to access financial and extension-advisory services such as weather, market, and agriculture advice. The use of remote sensing technologies which is one of the enabling technologies of the internet of things is also available within the sector but mostly accessible to researchers and agribusinesses. These ICTs are mostly used for research-related purposes. Nonetheless, usage and accessibility are still constrained by poor infrastructure and policy environment, fragmentation and low coordination in the agricultural research system, and low ICT skills and competencies of farmers.

The study also identified various theories underpinning the primary studies. The theoretical approaches tend to be narrowly focussed, identifying specific constructs that affect the adoption of technology and then using empirical evidence to demonstrate the robustness of the identified constructs. Several gaps were identified in the various theories. The TAM model for instance focus on attributes such as ease of use and perceived usefulness in determining factors that affect the adoption of technology. This theory tends to focus on the technical feasibility of the technology without taking into consideration other social, and legal factors that might impact adoption. Similarly, the DOI also falls in line with the proposed construct in the TAM, thereby lacking a social-cultural context. The SNT, however, discusses only social dimensions that affect adoption. This confirms Zewge and Dittrich (2017) study where appropriate design solutions that take social, as well as technical issues into account, are lacking in most literature. A comprehensive understanding of all the factors that impact adoption is, however important to prevent a mismatch between deployed technologies and the ecosystem of the local community. It was also evident that literature on frameworks for guiding the development of new ICT solutions is lacking as most of these theories focus on assessing the feasibility of already existing ICT solutions.

Regarding threats to the validity of the study, a review protocol was adapted in the study which helped in ensuring a rigorous review. In ensuring that an adequate number of relevant studies are retrieved, a wide range of databases was searched and both automatic and manual search approach was employed. We were, however, unable to exhaust all available databases due to inaccessible databases, making it a possibility that some relevant studies were missed. Also, pilot searches on search engines of selected electronic databases were conducted before constructing the keyword list. This helped ensure that the keywords used were related to the research topic and were able to retrieve all relevant studies of interest. Threats to publication bias were covered by including only published papers. Unpublished studies like information from company's website were excluded from the studies. This exclusion might, however, exclude other relevant studies leading to the introduction of bias. Data extraction form was also developed and the necessary fields that answer the research questions were outlined in consultation with fellow authors which helped in ensuring a detailed data extraction. In ensuring that bias at the reporting phase is reduced, evaluation by fellow authors was highly valued. The findings and conclusions were evaluated by individual authors and areas for improvements were identified and improved.

## 5 Conclusion

In this study, we have provided a systematic review of the state of art of ICT's within the agriculture sector in Africa. The results of the study will contribute to literature on ICT adoption in Africa. The review followed a detailed protocol and included primary studies from 2010 to 2019. We could identify 779 papers after applying our search string and 23 papers relevant to our research questions. The analysis of the primary studies revealed mobile-based services and platforms as the predominant ICT's used within the agriculture sector in Africa. Applications and services on mobile phones allow farmers to access extension-advisory services such as weather and market price information. The use of radios is still widely used in disseminating agriculture information to rural farmers. Several challenges that were found to impede the adoption of ICT include poor policy environment, low capacity, and poor technological infrastructure within the continent. Also, the theoretical frameworks identified from the primary studies provided theories for understanding and predicting the adoption of existing ICT technologies. The theories were predominantly focussed on assessing the technical feasibility of already existing technologies using identified constructs. Few of the theories took into account the social and cultural dimensions of the local context. This study, therefore, recommends a more holistic framework for guiding the development of ICT initiatives. The study also recommends the training and empowerment of smallholder farmers to enhance their ability to interact with new agriculture technologies. There is also the need for the development of a favourable policy and business environment that favours the use of ICT's and other digital technologies. Strong commitment, trust, and collaborations are also needed among the different actors in the agriculture value chain.

# References

Ahoa, E., Kassahun, A., Tekinerdogan,B. Business processes and information systems in the Ghana cocoa supply chain: A survey study, NJAS - Wageningen Journal of Life Sciences, Volume 92, 2020, 100323, ISSN 1573-5214, https://doi.org/10.1016/j.njas.2020.100323.

El Bilali, H., & Allahyari, M. S. (2018). Transition towards sustainability in agriculture and food systems: Role of information and communication technologies. *Information Processing in Agriculture*, *5*(4), 456–464.

Food and Agriculture Organisation ( FAO). (2017). Information and Communication Technology ( ICT) in Agriculture*: A Report to the G20 Agricultural Deputies*.

Iliyas, S. (2014). Impact of Information Technology in Agriculture Sector. *International Journal of Food, Agriculture and Veterinary Sciences*, *4*(2), 17-22.

Kassahun, A., Tekinerdogan, B. BITA*: Business-IT alignment framework of multiple collaborating organisations, Elsevier Information and Software Technology, Volume 127, 2020, 106345, ISSN 0950-5849, https://doi.org/10.1016/j.infsof.2020.106345.

Kitchenham, B., Pearl Brereton, O., Budgen, D., Turner, M., Bailey, J., & Linkman, S. (2009). Systematic literature reviews in software engineering - A systematic literature review. *Information and Software Technology*, *51***(**1), 7–15.

Köksal, Ö., Tekinerdogan, B. Architecture design approach for IoT-based farm management information systems. Precision Agric 20, 926–958 (2019). https://doi.org/10.1007/s11119-018-09624-8

Lwoga, E. T., & Sangeda, R. Z. (2019). ICTs and development in developing countries: A systematic review of reviews. *Electronic Journal of Information Systems in Developing Countries*, *85*(1), 1–17.

Moyo, J. M., Bah, E.-H. M., & Verdier-Chouchane, A. (2015). Transforming Africa's Agriculture to Improve Competitiveness. *The Africa Competitiveness Report.* 37–52.

OECD (2017), Science, technology and industry scoreboard 2017. The digital transformation

OECD Publishing, Paris, France. Available at https://doi.org/10.1787/20725345.

Technical Centre for Agriculture and Rural Cooperation (CTA). (2019). The Digitalization of African Agriculture Report.

Verdouw, C., Sundmaeker, H., Tekinerdogan, B., Conzon,D., Montanaro,T. Architecture framework of IoT-based food and farm systems: A multiple case study, Elsevier Computers and Electronics in Agriculture, Volume 165, 2019, ISSN 0168-1699, https://doi.org/10.1016/j.compag.2019.104939.

Wolfert, S., Ge, L., Verdouw, C., & Bogaardt, J.M (2017). Big Data in Smart Farming – A review. *Agricultural Systems*, *153*, 69–80

World Bank. (2008). Agriculture Development. In *World Development Report, Agriculture for Development*.

Zhang, N., Wang, M., & Wang, N. (2002) Precision agriculture—a worldwide overview. *Computers and Electronics in Agriculture, 36*, 113-132.

Zewge, A., & Dittrich, Y. (2017). Systematic mapping study of information technology for development in agriculture (the case of developing countries). *Electronic Journal of Information Systems in Developing Countries*, *82*(1), 1–25.

**Appendix A: List of primary studies**

1. Owusu, A.B., Yankson, P.W.K., & Frimpong, S. (2017) Smallholder farmers' knowledge of mobile telephone use: Gender perspectives and implications for agricultural market development. Progress in Development Studies, 18, 36-51.

2. Hudson, H.E., Leclair, M., Pelletier, B., & Sullivan, B. (2017). Using radio and interactive ICTs to improve food security among smallholder farmers in Sub-Saharan Africa. Telecommunications Policy, 4, 670-684.

3. Kante, M., Oboko, R., & Chepken, C. (2017). Influence of perception and quality of ICT-based agricultural input information on use of ICTs by farmers in developing countries: Case of Sikasso in Mali. Electronic Journal of Information Systems in Developing Countries 83(9), 1-21.

4. Barakabitze, A.A., Fue, K.G., & Sanga, C.A. (2017). The use of participatory approaches in developing ICT-based systems for disseminating agricultural knowledge and information for farmers in developing countries: The case of Tanzania. Electronic Journal of Information Systems in Developing Countries 78(8),1-23.

5. Mugwisi, T., Mostert J., & Ocholla, D.N. (2015). Access to and Utilization of Information and Communication Technologies by Agricultural Researchers and Extension Workers in Zimbabwe. Information Technology for Development, 21(1), 67-84.

6. Mwombe, S.O.L., Mugivane, F.I., Adolwa, I.S., & Nderitu, J.H. (2014) Evaluation of Information and Communication Technology Utilization by Small Holder Banana Farmers in Gatanga District, Kenya. The Journal of Agricultural Education and Extension, 20(2), 247-26.,

7. Aleke, B., Ojiako, U., & Wainwright, D.W. (2011). ICT adoption in developing countries: Perspectives from small-scale agribusinesses. Journal of Enterprise Information Management, 24 (1), 68-84.

8. Misaki, E., Apiola, M., Gaiani, S. (2016). Technology for small scale farmers in Tanzania: A design science research approach. Electronic Journal of Information Systems in Developing Countries, 74(4), 1-15.

9. Barakabitze, A.A., Kitindi, E.J., Sanga, C., Shabani, A., Philipo, J., & Kibirige, G. (2015). New technologies for disseminating and communicating agriculture knowledge and information: Challenges for agricultural research institutes in Tanzania. Electronic Journal of Information Systems in Developing Countries, 70 (2), 1-22.

10. Kabbiri, R., Dora, M., Kumar, V., Elepue, G., & Gellyncka, X. (2018). Mobile phone adoption in agri-food sector: Are farmers in Sub-Saharan African connected? Technological Forecasting & Social Change, 131, 253–261.

11. Jere, N.J., & Maharaj, M. S. (2017). Evaluating the influence of information and communications technology on food security. South African Journal of Information Management, 19(1), a745.

12. Beza, E., Reidsma, P., Poortvliet, P.M., Belay, M.M., & Bijen, B.S. (2018). Exploring farmers' intentions to adopt mobile Short Message Service (SMS) for citizen science in agriculture. Computers and Electronics in Agriculture, 151, 295–310.

13. Wyche, S., & Steinfield, C. (2016). Why Don't Farmers Use Cell Phones to Access Market Prices? Technology Affordances and Barriers to Market Information Services Adoption in Rural Kenya. Information Technology for Development, 22(2), 320-333.

14. Kante, M., Oboko, R., Chepken C. (2018). An ICT model for increased adoption of farm input information in developing countries: A case in Sikasso, Mali. Information Processing in Agriculture, 6 (1), 26-46.

15. Mtega, W. P., & Msungu, A. C. (2013). Using Information and Communication Technologies for Enhancing the Accessibility of Agricultural Information for Improved Agricultural Production in Tanzania. Electronic Journal of Information Systems in Developing Countries, 56(1), 1-14.

16. Kiambi, D. (2018). The use of Information Communication and Technology in advancement of African agriculture. African Journal of Agricultural Research, 13(39), 2025-2036.

17. Meijer, S.S., Catacutan, D., Ajayi, C.O., Sileshi, G. W., & Nieuwenhuis, M. (2015) The role of knowledge, attitudes, and perceptions in the uptake of agricultural and agroforestry innovations among smallholder farmers in sub-Saharan Africa. International Journal of Agricultural Sustainability, 13(1), 40-54.

18. Maredia, M.K., Reyes, B., Ba, M.N., Dabire, C.L., Pittendrigh, B., & Bravo, J. B. (2017). Can mobile phone-based animated videos induce learning and technology adoption among low literate farmers? A field experiment in Burkina Faso. Information Technology for Development, 24 (3), 429-460.

19. Mubichi, F., & Freeman, K. (2017). ICT use by smallholder farmers in rural Mozambique: a case study of two villages in central Mozambique. Journal of Rural Social Sciences, 32(2), 1–19.

20. Aleke, B., Ojiako, U., & Wainwright, D. (2011). Social networks among small agribusiness in Nigeria. Society and Business Review, 6 (3), 214-228.

21. Otene, V.A., Ezihe, J.A.C., & Torgenga, F. S. (2018). Assessment of Mobile Phone Usage Among Farmers in Keana Local Government Area of Nasarawa State, Nigeria. Journal of Agricultural & Food Information, 19(2), 141-148.

22. Magesa, M.M., Michael, K., & Ko, J. (2017). Towards a framework for accessing agricultural market information. Electronic Journal of Information Systems in Developing Countries, 66 (3), 1-16.

23. Awuor, F., Raburu, G., Onditi, A., & Rambim, D. (2016). Building E-Agriculture Framework in Kenya. Journal of Agricultural Informatics, 7 (1), 75-93.

# Appendix B: Quality Assessment Checklist

| Study | Quality of Reporting | | | Rigour | | | Relevance | | Credibility | | Quality of reporting | Rigour | Relevance | Credibility | Total |
|---|---|---|---|---|---|---|---|---|---|---|---|---|---|---|---|
| | Q1 | Q2 | Q3 | Q4 | Q5 | Q6 | Q7 | Q8 | Q9 | Q10 | | | | | |
| 1 | 1 | 1 | 1 | 0 | 1 | 1 | 1 | 0.5 | 1 | 0 | 3 | 2 | 1.5 | 1 | 7.5 |
| 2 | 1 | 1 | 0.5 | 0 | 0.5 | 1 | 1 | 1 | 1 | 0 | 2.5 | 1.5 | 2 | 1 | 7 |
| 3 | 1 | 1 | 1 | 1 | 1 | 1 | 1 | 0.5 | 1 | 0.5 | 3 | 3 | 1.5 | 1.5 | 9 |
| 4 | 1 | 1 | 1 | 0.5 | 1 | 1 | 1 | 1 | 1 | 0.5 | 3 | 2.5 | 2 | 1.5 | 9 |
| 5 | 1 | 0.5 | 1 | 1 | 0.5 | 1 | 0.5 | 0.5 | 1 | 0 | 2.5 | 2.5 | 1 | 1 | 7 |
| 6 | 1 | 1 | 1 | 1 | 1 | 1 | 0.5 | 1 | 1 | 0 | 3 | 3 | 1.5 | 1 | 8.5 |
| 7 | 1 | 1 | 1 | 0.5 | 1 | 0.5 | 1 | 0.5 | 1 | 1 | 3 | 2 | 1.5 | 2 | 8.5 |
| 8 | 1 | 1 | 1 | 0 | 1 | 1 | 0.5 | 0.5 | 1 | 0.5 | 3 | 2 | 1 | 1.5 | 7.5 |
| 9 | 1 | 1 | 1 | 0.5 | 1 | 1 | 1 | 0.5 | 1 | 1 | 3 | 2.5 | 1.5 | 2 | 9 |
| 10 | 1 | 1 | 0.5 | 1 | 0.5 | 0.5 | 0.5 | 1 | 1 | 1 | 2.5 | 2 | 1.5 | 2 | 8 |
| 11 | 1 | 1 | 0.5 | 1 | 0.5 | 1 | 0.5 | 0.5 | 1 | 0.5 | 2.5 | 2.5 | 1 | 1.5 | 7.5 |
| 12 | 1 | 1 | 1 | 1 | 0.5 | 0.5 | 0.5 | 1 | 1 | 1 | 3 | 2 | 1.5 | 2 | 8.5 |
| 13 | 0.5 | 1 | 1 | 0.5 | 1 | 1 | 1 | 1 | 1 | 1 | 2.5 | 2.5 | 2 | 2 | 9 |
| 14 | 1 | 0.5 | 1 | 1 | 0.5 | 1 | 1 | 1 | 1 | 0.5 | 2.5 | 2.5 | 2 | 1.5 | 8.5 |
| 15 | 0.5 | 1 | 1 | 0.5 | 1 | 1 | 0.5 | 1 | 1 | 0 | 2.5 | 2.5 | 1.5 | 1 | 7.5 |
| 16 | 1 | 1 | 0.5 | 1 | 0.5 | 1 | 1 | 0.5 | 1 | 0 | 2.5 | 2.5 | 1.5 | 1 | 7.5 |
| 17 | 1 | 1 | 1 | 1 | 0.5 | 1 | 1 | 0.5 | 1 | 0 | 3 | 2.5 | 1.5 | 1 | 8 |
| 18 | 1 | 1 | 1 | 0.5 | 1 | 0.5 | 1 | 1 | 1 | 0.5 | 3 | 2 | 2 | 1.5 | 8.5 |
| 19 | 1 | 1 | 1 | 0.5 | 1 | 1 | 1 | 0.5 | 1 | 0 | 3 | 2.5 | 1.5 | 1 | 8 |
| 20 | 1 | 1 | 0.5 | 1 | 1 | 0.5 | 1 | 0.5 | 1 | 1 | 2.5 | 2.5 | 1.5 | 2 | 8.5 |
| 21 | 1 | 1 | 0.5 | 0 | 1 | 1 | 1 | 1 | 1 | 0 | 2.5 | 2 | 2 | 1 | 7.5 |
| 22 | 1 | 0.5 | 1 | 0.5 | 0 | 1 | 0.5 | 0.5 | 1 | 0.5 | 2.5 | 1.5 | 1 | 1.5 | 6.5 |
| 23 | 1 | 1 | 0.5 | 0.5 | 0.5 | 1 | 0.5 | 0.5 | 1 | 0 | 2.5 | 2 | 1 | 1 | 6.5 |

# Appendix C: Data Extraction Form

| Extracted Element | Contents |
|---|---|
| *General Information* | |
| ID | Unique ID for the study |
| Authors | |
| Title | Full title of the paper |
| Year | The publication year |
| Source Title | The Publication channel |
| Document Type | + Journal  +Article |
| Repository | Scopus, ScienceDirect, Wiley Online, Abi/Inform, Taylor and Francis |
| *Study Description* | |
| Study design | + Survey   + Case study   + Experiment |
| Considered ICT domain | |
| Unit of analysis | Targeted agriculture stakeholders |
| Country Scope | |
| Considered Agriculture domain | |
| Identified Theoretical framework | +Yes    +No |
| Identified Challenges | + Yes    +No |
| *Evaluation* | |
| Quality Assessment | Quality, Rigour, Relevance, Credibility |